\documentclass[]{raa}            

\usepackage{graphicx,times}             

\begin{document}

\title{New variable Stars from the Photographic Archive: Semi-automated Discoveries,
Attempts of Automatic Classification, and the New Field 104 Her}

   \volnopage{Vol.0 (200x) No.0, 000--000}      
   \setcounter{page}{1}          

   \author{S. V. Antipin
      \inst{1}
   \and I. Becker
      \inst{2}
   \and A. A. Belinski
      \inst{1}
   \and D. M. Kolesnikova
      \inst{3}
   \and K. Pichara
      \inst{4,2}
   \and N. N. Samus
      \inst{3,1}
   \and K. V. Sokolovsky
      \inst{5,1,7}
   \and A. V. Zharova
      \inst{1}
   \and A. M. Zubareva
      \inst{3,1}
   }

\institute{P.K. Sternberg Astronomical Institute, M.V. Lomonosov
Moscow University, 13, University Ave., Moscow 119234,
Russia; {\it serge\_ant@inbox.ru} \\
        \and
        Pontificia Universidad Cat\'olica de Chile, Santiago, Chile\\
        \and
        Institute of Astronomy, Russian Academy of Sciences, Moscow, 119017, Russia \\
        \and
        Institute for Applied Computational Science, Harvard University,
        Cambridge, MA, USA
        \and
        IAASARS, National Observatory of Athens, 15236 Penteli, Greece
        \and
Institute of Astronomy, Russian Academy of Sciences, 48,
Pyatnitskaya Str., Moscow 109017, Russia\\
        \and
        Astro Space Center of Lebedev Physical Institute, Profsoyuznaya
Str. 84/32, 117997 Moscow, Russia
   }

\date{Received~~2017 November 24; accepted~~2017~~month day}

\abstract{Using 172 plates taken with the 40-cm astrograph of the
Sternberg Astronomical Institute (Lomonosov Moscow University) in
1976--1994 and digitized with the resolution of 2400 dpi, we
discovered and studied 275 new variable stars. We present the list
of our new variables with all necessary information concerning
their brightness variations. As in our earlier studies, the new
discoveries show a rather large number of high-amplitude Delta
Scuti variables, predicting that many stars of this type remain
not detected in the whole sky. We also performed automated
classification of the newly discovered variable stars based on the
Random Forest algorithm. The results of the automated
classification were compared to traditional classification and
showed that automated classification was possible even with noisy
photographic data. However, further improvement of automated
techniques is needed, which is especially important having in mind
the very large numbers of new discoveries expected from all-sky
surveys. }

   \authorrunning{S. V. Antipin et al.}            
   \titlerunning{New Variable Stars from the Photographic Archive}  

   \maketitle

%
%
\section{Introduction}           
\label{sect:intro}

The 20th century (more accurately, the time interval between 1880s
and 1990s) was the era of astronomical photography on plates and
films. The number of analog photographs at observatories of the
world is estimated to be about two millions. Information on the
existing plate stacks is collected in the Wide-field Plate
Database (www.wfpdb.org), founded by M.K.~Tsvetkov (Bulgaria).

Digitizing plate archives permits users, at the observatories
where the archives are stored and elsewhere, to use this vast
information applying modern digital reduction techniques for
scientific studies of importance for today's astronomy. It also
provides safety against events capable to destroy the analog
photographs: fires, floods, etc., like the water-main break at
Harvard Observatory on January~16, 2016. Harvard plate stacks, the
world-largest collection where about 500000 sky photographs are
kept, fortunately survived the terrible accident due to
effectiveness of emergency teams, but it is impossible to predict
what accident happens next and where. The process of digitizing
Harvard stacks was started long before the accident and is under
way quite successfully (e.g., Grindlay et al. 2009;
dasch.rc.fas.harvard.edu/index.php).

At Moscow Observatory (now  Sternberg Astronomical Institute of
Lomonosov Moscow University, SAI), direct photographs of the
starry sky were taken regularly in 1895--1995 using different
telescopes, with diameters of their objectives from 10 to 70~cm,
focal lengths from 64 to 1050~cm, installed at SAI sites in
Moscow, Crimea, and elsewhere. The founder of the Moscow plate
stacks was Prof. S.N. Blazhko (1870--1956). Details about this
plate collection can be found in Shugarov et al. (1999). The total
number of sky photographs (direct images) in the Moscow stacks is
estimated as 60000, they were used in different SAI departments
for different purposes.

The most important part of the Moscow plate stacks are 22300
plates taken with the 40-cm astrograph ($f = 160$~cm,
$10^\circ\times10^\circ$ field of view). This wide-field
multi-lens astrograph was initially ordered by C. Hoffmeister
(1892--1968) for Sonneberg Observatory (Germany). In 1938--1945,
1658 sky photographs were obtained with the astrograph. In 1945,
it  was selected as a part of war reparations by B.V. Kukarkin
(1909--1977), at that time an officer of the Soviet Army, later
one of the founders of the General Catalogue of Variable Stars. In
July, 1948 -- April, 1951 the telescope was in operation at Simeiz
Observatory, Crimea; in June, 1951 -- February, 1958, at Kuchino
Observatory near Moscow; in May, 1958, it became the first
instrument in operation at the newly established Crimean station
of the SAI in Nauchny settlement, Crimea (near the new territory
of the Crimean Astrophysical Observatory). At this site, more than
20800 plates were taken. The limiting magnitude in the central
parts of the field, for our typical 45-minute exposure time, is
$17^m-18^mV$. Unfortunately, distortions in the plate corners are
large, deteriorating the limiting magnitude.

The main purpose of star photographs taken with the 40-cm
astrograph were studies of variable stars. The program of
observations with the astrograph included many fields. Rich fields
contain about 300--500 sky photographs. In cases of overlapping
fields, a star can sometimes be found on 700 plates or even more.

We describe our work on digitizing the Moscow plate collection in
Section~2. Section~3 presents our new variable-star discoveries.
Section~4 deals with statistics of HADS variable stars among
variable stars discovered by us. In section~5, we discuss an
attempt of automated classification of the new variable stars
found in this field. Our conclusions are summarized in Section~6.

\section{Digitizing Moscow Plates}

Scanning plates of Moscow collection was started in 2006 using two
CREO EverSmart Supreme scanners, with resolution about 2500~dpi.
Unfortunately, electronics of these scanners became out of order
several years later and could not be repaired. In 2013, following
advice from M.K. Tsvetkov, we purchased an Epson Expression
11000XL scanner and continued digitizing our collection using the
resolution of 2400~dpi. Due to requirements of funding in Russia's
science, we decided to arrange the work so that scientific results
would not be delayed till completion of the scanning proper. After
finishing digitizing a field, we search for variable stars using
the scans and study the discovered variables.

When searching for variable stars, determining their periods,
plotting light curves, we widely use the VaST software package
developed by Sokolovsky and Lebedev (2017). VaST relies on
SExtractor (Bertin and Arnouts 1996) for detecting sources,
measuring their positions (in pixel coordinates) and brightness.

We apply the following procedure to extract light curves from a
series of digitized photographic images of a given sky region.

The grayscale TIFF images produced by the scanner are converted to
the FITS format using a custom-made code
(ftp://scan.sai.msu.ru/pub/software/tiff2fits/) and the
mid-exposure Julian date (extracted from telescope logbooks) is
recorded in the FITS header.

Pixel coordinates of a bright star visible in all images are used
as the reference point to cut each image into overlapping
$1.2^\circ\times1.2^\circ$ subfields. The subfields are positioned
relative to the reference star and cover the same sky area even
for plates that have a large offset from the nominal center of the
field. We neglect the plate rotation at this stage, as it is known
not to exceed $3^\circ$ for any of the plates digitized so far.

Then, SExtractor (Bertin and Arnouts 1996) is applied to each
subfield image to perform source extraction and circular aperture
photometry. The VaST code (Sokolovsky and Lebedev 2017) performs
cross-matching of the source lists (as the subfield images are not
aligned perfectly) and constructs light curves in an instrumental
magnitude scale. As the photographic density is a non-linear
function of the number of incoming photons, we use the relation
suggested by Bacher et al. (2005) to put the measured instrumental
magnitudes on the absolute scale set by the B magnitudes of
USNO-B1.0 (Monet et al. 2003) stars within each subfield.

Astrometry.net (Lang et al. 2010; Hogg et al. 2008) is used to
perform blind plate solution for each subfield needed to extract
equatorial coordinates of the detected objects.

New variables discovered in the course of our project are given
provisional variable-star names in the specially introduced MDV
(Moscow Digital Variable) series. Test experiments with scanning
small parts of plates resulted in discovering MDV1-- MDV 38. The
list of these stars can be found in Kolesnikova et al. (2010). The
first $10^\circ\times10^\circ$ field of the 40-cm astrograph
completely scanned and searched for variable stars by us was that
centered at the star 66 Oph ($18^h00.3^m$, $+04^\circ22'$,
J2000.0). In this field, we discovered and studied 480 new
variable stars, MDV39--MDV518 (Kolesnikova et al. 2008, 2010). The
77 new variable stars discovered in the field of SA9 (the center
at $03^h11.5^m$, $+60^\circ38'$, J2000.0; MDV519--MDV595) are
discussed in Sokolovsky et al. (2014). In the present paper, we
present our new results obtained in the field centered at the star
104 Her ($18^h11.9^m$, $+31^\circ24'$, J2000.0). We have
discovered and studied 275 new variable stars.

\section{New Variable Stars}

The plate stacks of the SAI contain 172 plates centered at the
star 104 Her ($l=58^\circ$, $b=22^\circ$; 1976 March 31 -- 1994
October 8). This is a moderately high-latitude field, and the star
density is not very high. The plates were scanned with the Epson
Expression 11000XL scanner, the resolution being 2400~dpi, and
searched for new variable stars using the VaST software.
Identified variable stars were checked in the GCVS and VSX
databases for being new. If photographic data, which have
comparatively low photometric accuracy, left doubt whether a
particular star was variable or not, we checked it using
additional sources of photometric information: the ROTSE/NSVS
database (Wo\'zniak et al. 2004a), Catalina database (Drake et al.
2009), and ASAS-SN database (Kochanek et al. 2017). Six stars in
the final list were confirmed on the base of specially arranged
CCD observations using telescopes of the SAI Crimean laboratory.

Figure 1 explains our technique of photographic photometry and
search for variable stars. The VaST code checks all stars
deviating along the ordinate from the most densely populated cloud
for periodicity; additionally, non-periodic, strongly deviating
stars (that can be non-periodic variables) are inspected by eye.
Some of the outliers, those not marked with numbers, are mainly
blended stars.

\begin{figure}
   \centering
   \includegraphics[width=13cm, angle=0]{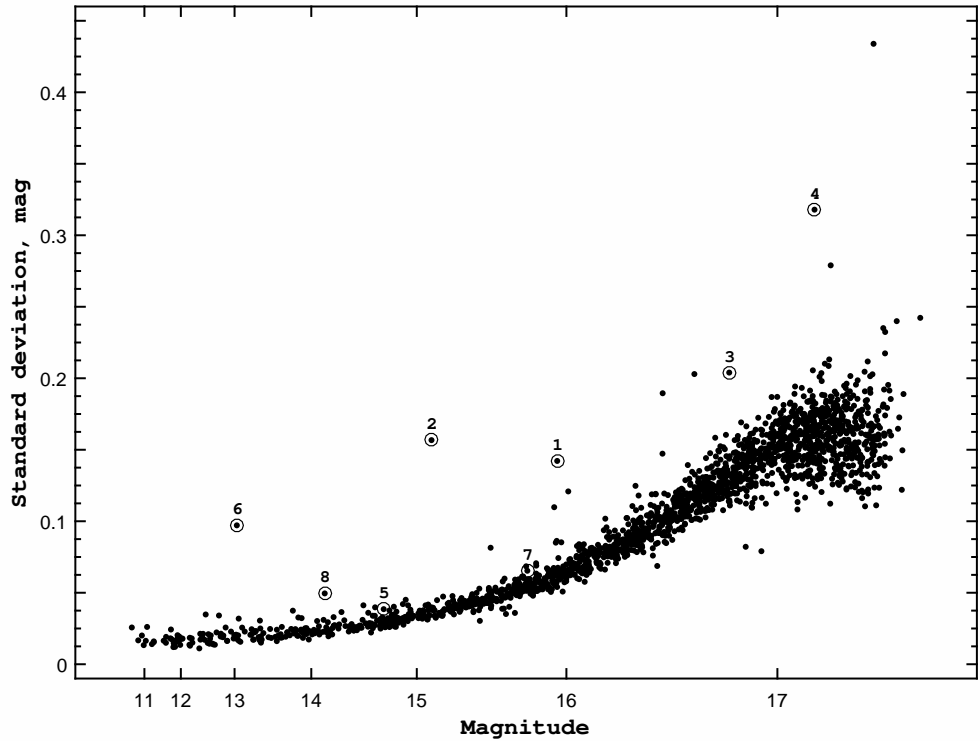}
   \caption{The diagram explaining the process of our search for variable stars.
   The circled stars are new and known variable stars. 1: NSVS J1812081+302718 (Wo\'zniak et al. 2004b);
   2: V1335 Her; 3: MDV 684; 4: MDV 700; 5: MDV 701; 6: MDV 705;
   7:  MDV 710; 8:  MDV 713.}
   \label{Fig0}
   \end{figure}

Our final list of newly discovered variable stars is presented in
Table 1. It contains 275 new variables (MDV596--MDV870).

The columns of the Table present: the MDV number of the variable;
its equatorial coordinates (J2000.0); its variability type; its
period (for periodic variables); its variability range
(photographic B magnitudes at maximum and minimum light, also in
the secondary minimum for eclipsing stars); number of the remark
below the table. The light curves of the variables can be found at
our web site (http://www.sai.msu.su/gcvs/digit/digit.html).

We performed classification of all the new discoveries using the
traditional approach used in the team of the General Catalogue of
Variable Stars some of the authors belong to. This classification
is principally based on the light curve shape but also takes into
account all available additional information (period, amplitude,
color index, etc.). Judging from our preferred classifications
(the third column of Table 1), more than a half of all discoveries
(150 stars) are eclipsing variables. Among pulsating variables in
Table 1, we often meet RR Lyrae variable stars (70 objects),
semiregular (SR) or irregular (LB, L) stars. There are also eight
high-amplitude Delta Scuti (HADS) variables, three Cepheids of the
second type (CW), and a variable of the RV Tauri (RVA) type. Two
variables are probable rotating spotted stars (type BY:). Figure 1
displays sample photographic light curves for variables of several
types. We present our CCD light curves for two stars in Fig.~2.

It is very interesting that our photographic search for new
variable stars made it possible to detect also low-amplitude
variable stars. The lowest peak-to-peak amplitude, $0.1^m$, was
found for the EW eclipsing star MDV640, quite reliably confirmed
with data from the ASAS-SN survey.

We will address the period distribution of newly discovered
eclipsing variable stars elsewhere. Here we only mention that, for
EW stars, this distribution is evidently shifted towards shorter
periods, as already noted by Kolesnikova et al. (2010) for our
discoveries in the field of 66 Oph.

\section{Statistics of HADS Variables}

Among short-period pulsating variable stars, a large fraction are
Delta Scuti variables, mainly located in the region of
intersection of the Cepheid instability strip with the main
sequence. Most of them have small pulsation amplitudes; however,
there exist Delta Scuti stars with $V$-band or photographic
amplitudes of $0.2^m$ or higher. In most studies, such stars are
called HADS (High Amplitude Delta Scuti) variables. This is not a
homogeneous variability type but a mixture of younger
(Population~I) stars and SX~Phoenicis variables that belong to old
galactic populations. A clear distinction between these
variability types has not yet been achieved (see Balona 2016). In
the following, we call all large-amplitude, short-period pulsating
variable stars HADS variables. Our data do not permit a study of
their multiperiodic behavior, quite typical of some of them.

From scans of the field centered at 66 Oph, we found 11 HADS stars
with peak-to-peak amplitudes of $0.2^m$ or higher. Kolesnikova et
al. (2010) remarked that this result permitted to expect the
number of HADS variables missing from the existing variable-star
catalogs to be quite large, the studied field's area being only
0.24\% of the whole area of the sky. The amplitudes of the new
HADS variables are not small, they can be easily discovered using
photographic plate collections and traditional techniques (eye
estimates or measurements with microphotometers).

Expectations for many new HADS variables were not quite fulfilled
in the field SA9 (Sokolovsky et al. 2014), where we discovered
only one HADS variable.

In the present study, we found eight new HADS variable stars (two
of them uncertain);  there are also doubts if MDV790 is an EW or a
HADS star, making the total number of new HADS stars nine (three
of them uncertain).

Thus, the total number  of HADS discoveries in the field of 300
square degrees (0.73\% of the sphere) is from 18 to 21, resulting
in an estimate of 2500--3000 for the expected number of HADS stars
for the whole sky. This estimate should be additionally somewhat
increased because of some variables remaining undiscovered in the
corners of the fields, where the 40-cm astrograph has very large
distortions.

As of November, 2017, the electronic version of the 5th edition of
the General Catalogue of Variable Stars (GCVS V; Samus et al.
2017) contains 212 stars that can be considered HADS variables
(note that the GCVS classification system does not explicitly
contain the HADS type, and we had to check classification
ourselves). Thus, very many HADS stars remain not included into
the GCVS. At first glance, it seems strange: as mentioned above,
we are discussing stars with variation amplitudes not too low.
However, traditional approaches to discoveries of variable stars,
their study, and their classification could, indeed, fail to find
HADS stars because of their short periods. Such periods are no
problem for modern computers, but they are not too easy to
determine without a computer, as it was done in early studies.

\newpage

{\small

\begin{table}
\begin{center}
\caption[]{New MDV variable stars in the field of 104 Her
}\label{Tab:publ-works}

 \begin{tabular}{lllllllll}
  \hline\noalign{\smallskip}
MDV & RA, Dec (J2000.0) & Type & Period& Max& Min &MinII& Epoch& Rem\\
  \hline\noalign{\smallskip}

596& 17:49:01.69 +31:23:16.2 &EB &   0.622119 & 14.30&  14.70& 14.60& min 2444110.340&  1,2\\
597& 17:49:39.63 +33:22:17.0 &RRAB&  0.572267 & 16.75&  17.80&      & max 2443671.404&  2\\
598& 17:50:28.69 +32:30:46.9 &SR  &54.0       &14.60 & 15.00 &       &max 2443697.3  &  1,3\\
599& 17:51:05.95 +30:03:06.8 &SR  & 24.5      & 13.40&  13.75&      & max 2444044.5  &  1\\
600& 17:53:04.73 +28:57:56.4 &EW  &  0.275626 & 16.20&  16.60& 16.40& min 2444105.367&  2\\
601& 17:53:23.23 +29:02:03.8 &EB  &  0.504457 & 15.75&  16.15& 15.90& min 2443757.332&  2\\
602& 17:54:24.12 +28:44:00.8 &SR  & 56.6:     & 15.10&  15.40&      & max 2443663.4  &  2,4\\
603& 17:54:36.03 +26:41:33.9 &EA  &  1.775815 & 15.05&  15.50&      & min 2443786.225&  2\\
604& 17:54:39.01 +28:27:49.5 &L:  &           & 11.90&  12.50&      &                &  5\\
605& 17:54:49.28 +26:44:32.8 &EA  &  1.22715  & 15.00&  15.35&      & min 2446357.25 &  2\\
606& 17:55:01.44 +30:36:54.2 &EB  &  0.806310 & 15.25&  15.50& 15.45& min 2443757.34 &  2\\
607& 17:55:03.07 +29:19:10.9 &RRAB&  0.557696 & 15.90&  16.95&      & max 2443430.232&  6\\
608& 17:55:33.64 +26:49:51.7 &EW  &  0.277426 & 15.65&  15.95& 15.90& min 2443196.416&  2\\
609& 17:55:52.95 +33:48:12.3 &RRC:&  0.198706 & 14.35&  14.65&      & max 2444873.229&  1,2,7\\
610& 17:55:53.33 +28:16:26.4 &RRC &  0.336777 & 16.30&  16.85&      & max 2444112.380&  2\\
611& 17:56:00.95 +28:59:47.3 &EA  &  1.673843 & 14.80&  15.50& 15.05& min 2443694.318&  2\\
612& 17:56:04.66 +33:01:22.9 &EA  &  6.31503  & 15.25&  15.85& 15.85& min 2444790.4  &  2,8\\
613& 17:56:42.28 +31:52:04.4 &LB  &           & 14.20&  14.70&      &                &  1\\
614& 17:56:57.55 +30:21:25.1 &EA  &  0.895829 & 15.55&  16.15& 15.75& min 2444456.31 &  2\\
615& 17:57:08.07 +32:54:23.6 &LB  &           & 11.30&  11.80&      &                &  1\\
616& 17:57:09.02 +30:46:03.4 &EW  &  0.282263 & 15.55&  15.80& 15.80& min 2445258.225&  2\\
617& 17:57:09.96 +28:34:17.8 &EA  &  2.76699  & 13.60&  13.90& 13.85& min 2442869.50 &  2,9,10\\
618& 17:57:36.38 +30:44:07.3 &BY: &  3.016    & 16.60&  17.10&\\
619& 17:57:43.74 +28:08:09.0 &EW  &  0.372968 & 16.45&  17.00& 16.90& min 2444073.419&  2\\
620& 17:57:46.06 +30:54:39.9 &LB  &           & 14.90&  15.20&      &                &  1\\
621& 17:57:49.15 +31:03:02.6 &EW  &  0.296551 & 16.70&  17.10& 17.00& min 2443672.320&  2\\
622& 17:58:12.27 +26:21:47.4 &EW  &  0.397835 & 15.95&  16.50& 16.50& min 2443701.418&  2\\
623& 17:58:56.10 +31:01:53.0 &EA  &  1.96592  & 14.85&  15.70& 14.95& min 2444162.26 &  2\\
624& 17:59:13.03 +27:11:23.6 &EW  &  0.359906 & 14.25&  14.40& 14.40& min 2445258.225&  1,2,11\\
625& 17:59:15.09 +30:14:55.5 &EW  &  0.341483 & 14.95&  15.30& 15.25& min 2444489.313&  2,12\\
626& 17:59:16.44 +33:06:46.5 &EW  &  0.470138 & 14.60&  14.90& 14.85& min 2444461.315&  2\\
627& 17:59:24.01 +27:45:57.0 &RRC &  0.286315 & 14.15&  14.35&      & max 2443199.553&  2,7\\
628& 17:59:49.75 +29:53:09.1 &EW  &  0.340187 & 16.05&  16.30& 16.25& min 2445228.280&  2\\
629& 18:00:10.98 +32:58:11.2 &RRC &  0.337218 & 17.10&  17.60&      & max 2445228.280&  2\\
630& 18:01:18.15 +27:35:04.7 &EW  &  0.359731 & 14.70&  15.00& 14.95& min 2443671.404&  2\\
631& 18:01:33.73 +34:35:43.6 &EA  &  0.866945 & 16.10&  16.85&      & min 2444055.447&  2\\
632& 18:02:07.99 +29:27:52.9 &EW  &  0.401450 & 14.95&  15.20& 15.15& min 2443670.382&  2\\
633& 18:02:11.32 +28:52:06.1 &EA  &  0.727779 & 16.10&  16.80& 16.15& min 2443199.553&  2\\
634& 18:02:34.43 +30:41:22.7 &RRAB&  0.614662 & 14.90&  15.50&      & max 2444823.293&  2,9\\
635& 18:02:57.31 +27:05:42.5 &EW  &  0.473822 & 16.10&  16.40& 16.35& min 2444728.514&  2\\
636& 18:03:35.03 +32:18:34.9 &RRAB&  0.646628 & 15.40&  16.90&      & max 2443691.411&  9,13\\
637& 18:03:48.03 +29:15:31.9 &EW  &  0.415536 & 15.50&  15.80& 15.75& min 2444107.363&  2\\
638& 18:04:16.05 +33:37:41.2 &RRC &  0.378115 & 15.40&  15.90&      & max 2447676.378&  2,14\\
639& 18:04:32.84 +29:39:43.0 &RRC &  0.333640 & 14.45&  14.60&      & max 2443787.219&  2\\

  \noalign{\smallskip}\hline
\end{tabular}
\end{center}
\end{table}
}

\newpage
{\small

\begin{table}
\begin{center}
\caption[]{Table 1(continued)
}\label{Tab:publ-works}

 \begin{tabular}{lllllllll}
  \hline\noalign{\smallskip}
MDV & RA, Dec (J2000.0) & Type & Period& Max& Min &MinII& Epoch& Rem\\
  \hline\noalign{\smallskip}
640 &18:04:39.39 +29:47:56.8 &EW   & 0.428358  &13.75  &13.85 &13.85 &min 2444463.332 & 15\\
641 &18:04:42.04 +30:41:36.3 &EA   &0.917043   &14.85  &15.30 &14.95 &min 2443695.411\\
642 &18:04:47.57 +28:14:17.8 &RRAB & 0.478659  &15.80  &16.90 &      &max 2443253.482 & 2\\
643 &18:04:49.89 +30:17:20.1 &RRC  & 0.313303  &16.45  &16.90 &      &max 2445258.225\\
644 &18:04:51.62 +30:01:29.9 &EW   & 0.383206  &15.65  &15.90 &15.85 &min 2444823.293 & 2\\
645 &18:04:55.83 +30:42:00.5 &EA   & 1.84559   &13.65  &13.85 &13.75 &min 2444046.39  & 1\\
646 &18:05:09.03 +29:05:24.7 &EA   & 3.0150    &13.10  &13.40 &13.20 &min 2443705.47  & 1\\
647 &18:05:12.49 +29:04:40.4 &EA   & 3.4917    &15.30  &15.60 &15.40 &min 2444105.37\\
648 &18:06:11.50 +34:38:51.4 &EW   & 0.523530  &14.85  &15.05 &15.00 &min 2444131.333&  2\\
649 &18:06:17.78 +34:37:48.7 &RRAB & 0.528967  &16.45  &17.30 &      &max 2444838.348& 2\\
650 &18:06:21.99 +32:32:35.5 &SR   &30.3:      &13.75  &14.05 &      &max 2443189.6  &  16\\
651 &18:06:22.76 +27:33:24.9 &EA   & 1.75983   &15.35  &16.00 &      &min 2444876.230&  2\\
652 &18:06:30.95 +26:59:30.5 &EW   & 0.359248  &15.85  &16.25 &16.15 &min 2448778.490&  17\\
653 &18:06:47.02 +28:27:17.1 &EW   & 0.452924  &14.15  &14.40 &14.35 &min 2444852.315&  1,2\\
654 &18:07:00.47 +27:32:46.5 &EB   & 0.838553  &13.95  &14.25 &14.05 &min 2443685.371&  1,2\\
655 &18:07:05.27 +30:58:47.8 &EW   & 0.242034  &15.20  &15.75 &15.70 &min 2444083.428&  2\\
656 &18:07:07.52 +32:17:44.0 &LB   &           &12.70  &13.05 &      &               &  1\\
657 &18:07:12.53 +32:54:22.9 &CWA: &15.07      &15.60  &15.90 &      &max 2443705.5  &  2\\
658 &18:07:24.82 +29:29:27.6 &EW   & 0.352261  &13.80  &14.00 &14.00 &min 2443691.373&  2\\
659 &18:07:25.09 +36:02:04.0 &EW   & 0.488014  &14.05  &14.50 &14.35 &min 2444818.416&  2\\
660 &18:07:27.49 +31:29:52.3 &EW:  & 0.485218  &16.60  &17.30 &      &min 2444875.237&  2\\
661 &18:07:36.34 +36:00:44.8 &EA   & 1.88865   &15.25  &15.75 &      &min 2444015.404&  2\\
662 &18:07:38.46 +30:46:10.6 &EW   & 0.255938  &16.20  &16.60 &16.60 &min 2443780.248&  2\\
663 &18:07:49.22 +32:57:16.4 &LB   &           &15.20  &16.00\\
664 &18:07:50.39 +29:17:21.2 &EW   & 0.595814  &15.20  &15.40 &15.40 &min 2443694.318&  2\\
665 &18:08:03.11 +28:53:38.4 &RRAB & 0.719960  &16.40  &17.10 &      &max 2444103.301&  2\\
666 &18:08:36.96 +33:46:27.7 &CWA: &16.8       &15.25  &15.50 &      &max 2443671.4  &  18\\
667 &18:08:56.57 +29:46:51.6 &RRAB & 0.462302  &16.20  &17.30 &      &max 2444109.291&  2\\
668 &18:09:20.36 +33:49:58.1 &LB   &           &14.90  &15.30 &      &               &  1\\
669 &18:09:42.19 +28:18:05.8 &RRAB & 0.538265  &16.20  &17.00 &      &max 2443253.482&  2\\
670 &18:10:23.71 +28:31:08.8 &EB   & 0.491929  &14.85  &15.10 &14.95 &min 2444110.340&  2\\
671 &18:10:36.63 +33:38:53.6 &EW   & 0.388084  &14.80  &14.95 &14.95 &min 2445205.333&  1,12\\
672 &18:10:37.86 +28:44:03.5 &RRAB & 0.665442  &15.40  &16.10 &      &max 2444815.421&  2\\
673 &18:10:43.76 +28:07:16.5 &EW   & 0.389823  &15.10  &15.35 &15.35 &min 2444162.263&  2\\
674 &18:10:54.53 +32:22:53.2 &HADS & 0.0847320 &16.40  &16.75 &      &max 2444876.230&  2\\
675 &18:11:03.37 +28:07:03.0 &EW   & 0.384549  &15.40  &15.70 &15.60 &min 2443282.418&  2\\
676 &18:11:07.44 +30:47:40.8 &EW   & 0.365738  &16.40  &16.90 &16.80 &min 2445258.225&  2\\
677 &18:11:07.68 +30:51:29.5 &EW   & 0.337050  &14.50  &14.70 &14.60 &min 2444053.409&  2\\
678 &18:11:10.54 +35:38:54.4 &RRC  & 0.322462  &16.40  &16.75 &      &max 2443282.418&  2\\
679 &18:11:20.05 +28:53:27.0 &RRAB & 0.633044  &16.15  &17.00 &      &max 2444810.340&  2\\
680 &18:11:31.04 +29:16:10.5 &SR   &41.9:      &13.85  &14.05 &      &               &  19\\
681 &18:11:32.19 +27:51:32.5 &SR  &172         &14.40  &14.70 &      &max 2443670    &  20\\
682 &18:11:33.12 +27:23:03.3 &EB   &22.9522    &15.00  &15.80 &15.20 &min 2444111.4  &  2\\
683 &18:11:52.18 +29:20:24.7 &RRC  & 0.416369  &15.25  &15.80 &      &max 2444075.403&  2\\
684 &18:11:53.37 +30:24:50.4 &RRAB & 0.715046  &16.50  &17.10 &      &max 2443686.417&  2\\

  \noalign{\smallskip}\hline
\end{tabular}
\end{center}
\end{table}
}

\newpage
{\small

\begin{table}
\begin{center}
\caption[]{Table 1(continued)
}\label{Tab:publ-works}

 \begin{tabular}{lllllllll}
  \hline\noalign{\smallskip}
MDV & RA, Dec (J2000.0) & Type & Period& Max& Min &MinII& Epoch& Rem\\
  \hline\noalign{\smallskip}

685& 18:12:12.49 +29:10:14.7 &EB  &  0.725712 & 14.70 & 15.10 &14.95& min 2444875.237&  2\\
686& 18:12:23.90 +33:50:05.7 &EB  &  0.465791 & 16.05 & 16.45 &16.20& min 2444880.345&  2\\
687& 18:12:24.30 +28:32:34.6 &EA  &  0.766385 & 15.60 & 16.10 &     & min 2444039.356&  2\\
688& 18:12:29.51 +26:36:01.3 &RRAB&  0.552409 & 15.80 & 16.80 &     & max 2443744.310&  2\\
689& 18:12:46.79 +27:31:43.9 &EW  &  0.297575 & 14.90 & 15.20 &15.10& min 2443723.388&\\
690& 18:12:52.34 +29:42:35.5 &RRAB&  0.549889 & 15.95 & 16.55 &     & max 2443253.482&  2\\
691& 18:12:56.78 +29:13:04.9 &RRAB&  0.565463 & 15.50 & 16.75 &     & max 2443717.377&  2\\
692& 18:12:58.04 +29:36:25.3 &EB  &  0.387342 & 15.55 & 15.80 &15.65& min 2443784.222&  2\\
693& 18:12:58.57 +28:59:28.2 &EB  &  0.741523 & 15.20 & 15.35 &15.25& min 2444873.229&\\
694& 18:13:18.86 +29:43:31.6 &RRAB&  0.565299 & 14.70 & 15.60 &     & max 2443672.327&  2\\
695& 18:13:20.08 +31:56:05.8 &SR  & 68.5:     & 12.80 & 13.15 &     & max 2444461.3  &  21\\
696& 18:13:33.92 +29:36:44.7 &LB  &           & 12.10 & 12.60 &     &                &  1\\
697& 18:13:39.28 +28:59:04.8 &SR  & 55.94     & 15.80 & 16.50 &     & max 2443744.3  &  22\\
698& 18:14:05.12 +33:50:57.5 &EB  &  0.697571 & 14.90 & 15.25 &15.00& min 2443722.368&  2\\
699& 18:14:05.89 +30:49:31.2 &RRAB&  0.470188 & 16.80 & 17.80 &     & max 2443699.457&  2\\
700& 18:14:15.07 +29:56:13.5 &RRAB&  0.628599 & 16.65 & 17.50 &     & max 2443663.483&  2\\
701& 18:14:15.68 +30:14:22.2 &EW  &  0.369758 & 14.65 & 14.80 &14.75& min 2443726.360&  15\\
702& 18:14:17.99 +27:10:07.6 &EA  &  3.384255 & 15.50 & 16.40 &15.60& min 2444397.450&\\
703& 18:14:18.07 +27:13:18.1 &EW  &  0.303709 & 15.60 & 16.00 &15.90& min 2444815.421&\\
704& 18:14:25.46 +27:22:53.6 &RRC:&  0.281205 & 16.15 & 16.60 &     & max 2443934.572&  7\\
705& 18:14:26.60 +30:11:44.3 &LB  &           & 11.80 & 13.40\\
706& 18:14:31.57 +34:01:47.1 &EW  &  0.326831 & 13.25 & 13.45 &13.45& min 2444459.318&  9\\
707& 18:14:38.47 +33:26:42.9 &EW  &  0.323163 & 14.10 & 14.30 &14.30& min 2443748.313&  2\\
708& 18:14:43.56 +27:28:39.9 &EW  &  0.309487 & 14.95 & 15.20 &15.20& min 2446357.247&  23\\
709& 18:14:43.90 +29:37:52.9 &RRC &  0.386866 & 16.05 & 16.50 &     & max 2442869.499&  24\\
710& 18:14:44.88 +29:58:29.4 &RRC:&  0.262310 & 15.70 & 15.85 &     & max 2444104.282&  7\\
711& 18:14:46.63 +29:21:05.3 &EW  &  0.272737 & 15.30 & 15.65 &15.60& min 2444104.282&  2\\
712& 18:15:02.50 +33:02:29.9 &EB  &  0.665691 & 15.55 & 15.80 &15.65& min 2444024.407&  2\\
713& 18:15:13.49 +29:51:56.2 &EA  &  0.885062 & 14.10 & 14.75 &     & min 2443717.377&  2\\
714& 18:15:19.69 +34:24:40.8 &EW  &  0.430040 & 13.55 & 13.75 &13.75& min 2443934.572&  2\\
715& 18:15:23.59 +28:14:46.9 &EW  &  0.368195 & 16.50 & 17.00 &16.85& min 2443759.336&  2\\
716& 18:15:34.17 +30:16:12.3 &EA  &  0.857063 & 15.95 & 16.60 &16.40& min 2444075.403&  2\\
717& 18:15:43.87 +28:17:03.1 &EB  &  0.545438 & 14.25 & 14.85 &14.40& min 2444085.315&  1,2\\
718& 18:15:44.02 +29:16:24.6 &EW  &  0.400190 & 15.90 & 16.40 &16.20& min 2444821.298&  2\\
719& 18:15:47.68 +28:25:15.8 &EW  &  0.385311 & 15.75 & 16.10 &16.05& min 2445228.280&  2\\
720& 18:16:00.71 +28:28:20.6 &RRAB&  0.555227 & 16.40 & 17.40 &     & max 2443691.411&  2\\
721& 18:16:02.36 +31:30:13.8 &EB: & 18.4546   & 15.40 & 15.60 &15.55& min 2444820.4\\
722& 18:16:22.12 +27:38:21.8 &RRAB&  0.504323 & 15.70 & 17.10 &     & max 2444838.348&  2\\
723& 18:16:28.80 +30:00:11.1 &RRAB&  0.550401 & 16.30 & 17.30 &     & max 2444818.416\\
724& 18:16:37.23 +29:32:33.9 &RRAB&  0.612923 & 16.15 & 16.90 &     & max 2444759.432\\
725& 18:16:44.90 +32:24:42.7 &HADS&  0.0761223& 16.05 & 16.30 &     & max 2443671.404&  2\\
726& 18:16:59.10 +27:25:36.5 &EW  &  0.408681 & 15.10 & 15.40 &15.30& min 2443701.418&  2\\
727& 18:17:20.88 +30:04:25.7 &EW  &  0.408577 & 15.30 & 15.50 &15.45& min 2444456.309\\
728& 18:17:30.78 +28:55:21.4 &RRAB&  0.533280 & 15.40 & 16.80 &     & max 2444046.393\\
729& 18:17:38.61 +28:30:23.2 &RRC &  0.288996 & 15.85 & 16.45 &     & max 2443815.196\\
730& 18:17:40.19 +30:59:23.9 &RRAB&  0.487686 & 15.80 & 17.00 &     & max 2443672.327&  25\\

  \noalign{\smallskip}\hline
\end{tabular}
\end{center}
\end{table}
}

\newpage
{\small

\begin{table}
\begin{center}
\caption[]{Table 1(continued)
}\label{Tab:publ-works}

 \begin{tabular}{lllllllll}
  \hline\noalign{\smallskip}
MDV & RA, Dec (J2000.0) & Type & Period& Max& Min &MinII& Epoch& Rem\\
  \hline\noalign{\smallskip}

731 &18:17:41.94 +30:46:44.9 &RRAB & 0.555318  &15.90 & 16.90 &      &max 2444458.316\\
732 &18:18:03.72 +27:03:02.9 &HADS & 0.115625  &15.65 & 16.05 &      &max 2444110.340&  15\\
733 &18:18:07.74 +33:26:11.9 &LB   &           &14.50 & 14.75 &      &               &  1\\
734 &18:18:08.52 +28:27:35.3 &EW   & 0.358719  &16.00 & 16.35 &16.35 &min 2444081.379\\
735 &18:18:12.30 +33:12:58.4 &EA   & 2.14430   &14.75 & 15.35 &      &min 2443754.341&  2\\
736 &18:18:24.28 +27:55:43.8 &EW   & 0.341269  &15.95 & 16.30 &16.30 &min 2444823.293&  2\\
737 &18:18:58.34 +27:11:56.6 &EW   & 0.300959  &15.35 & 15.60 &15.55 &min 2444781.425&  2\\
738 &18:19:02.75 +33:46:27.5 &SR   &51.8       &13.30 & 13.45 &      &max 2444846.4  &  26\\
739 &18:19:07.06 +28:33:07.2 &RRC  & 0.390775  &16.65 & 17.05 &      &max 2444111.411\\
740 &18:19:13.58 +28:17:35.2 &EW   & 0.303963  &14.55 & 14.70 &14.70 &min 2444815.421\\
741 &18:19:17.56 +28:52:44.5 &EW   & 0.375928  &14.50 & 14.90 &14.85 &min 2444109.291&  1\\
742 &18:19:19.38 +31:12:16.2 &EA   & 1.51602   &15.15 & 15.70 &      &min 2444397.450&  2\\
743 &18:19:22.77 +28:29:14.5 &SR   &29.0       &12.40 & 13.00 &      &max 24437123.4 &  27\\
744 &18:19:23.55 +29:10:58.1 &EW   & 0.332247  &15.55 & 15.80 &15.80 &min 2444847.395\\
745 &18:19:30.60 +26:25:51.5 &EW   & 0.318962  &15.80 & 16.30 &16.20 &min 2443686.327&  1,2\\
746 &18:19:32.62 +30:13:38.5 &RRAB & 0.833771  &16.45 & 17.05 &      &max 2445192.304\\
747 &18:19:32.83 +29:21:29.2 &EW   & 0.519793  &16.60 & 17.10 &17.00 &min 2444494.283&  28\\
748 &18:19:34.10 +30:15:58.0 &RRC  & 0.382189  &16.45 & 17.00 &      &max 2444494.283\\
749 &18:20:07.41 +32:33:43.0 &EB   &12.0577    &13.75 & 13.95 &13.90 &min 2444045.4  &  1\\
750 &18:20:15.45 +33:27:06.2 &EW   & 0.405780  &13.35 & 13.50 &13.45 &min 2443814.217&  1\\
751 &18:20:20.06 +34:11:24.5 &EW   & 0.405052  &14.10 & 14.30 &14.30 &min 2444995.653&  2\\
752 &18:20:22.49 +34:19:38.5 &EA   & 2.73621   &13.35 & 13.95 &      &min 2444104.282&  1\\
753 &18:20:23.30 +29:20:39.7 &RRAB & 0.454574  &15.00 & 16.30 &      &max 2444466.316\\
754 &18:20:30.58 +29:24:30.0 &EB   & 0.501325  &15.05 & 15.35 &15.20 &min 2444459.318\\
755 &18:20:37.05 +30:29:59.3 &LB   &           &11.90 & 12.70 &      &               &  1\\
756 &18:20:40.70 +32:16:29.5 &SR   &47.7       &15.35 & 15.70 &      &max 2443672.3  &  29\\
757 &18:20:50.48 +29:03:24.1 &EA   & 1.08921   &15.60 & 16.15 &15.70 &min 2443699.457\\
758 &18:20:56.61 +33:04:52.4 &RRAB & 0.590969  &15.75 & 16.45 &      &max 2444781.425\\
759 &18:20:58.93 +28:50:06.6 &EW   & 0.281518  &16.30 & 16.70 &16.60 &min 2444044.462\\
760 &18:21:08.37 +29:33:03.8 &EW   & 0.408780  &16.55 & 17.10 &16.90 &min 2444493.449\\
761 &18:21:11.26 +32:06:16.0 &EW   & 0.359348  &15.95 & 16.25 &16.20 &min 2444876.230\\
762 &18:21:20.39 +31:15:40.5 &EW   & 0.409723  &14.45 & 14.70 &14.65 &min 2444459.318\\
763 &18:21:21.91 +30:33:08.7 &RRAB & 0.436579  &13.90 & 15.45 &      &max 2443783.253\\
764 &18:21:22.79 +31:06:32.1 &EA   & 1.86917   &14.85 & 15.15 &15.00 &min 2444459.318\\
765 &18:21:45.49 +27:39:04.2 &EA   & 2.31034   &15.75 & 16.60 &15.85 &min 2443757.332\\
766 &18:21:50.88 +35:51:14.4 &SR   &204.7      &13.60 & 13.90 &      &max 2449267.2  &  29\\
767 &18:22:02.46 +29:45:42.7 &EW   & 0.388485  &14.90 & 15.15 &15.05 &min 2444039.356\\
768 &18:22:11.55 +32:18:50.4 &EW   & 0.377804  &15.15 & 15.40 &15.35 &min 2444461.315&  9\\
769 &18:22:12.20 +28:11:59.7 &RRC  & 0.319727  &14.55 & 14.75 &      &max 2444081.379\\
770 &18:22:27.30 +27:32:34.8 &CWA: &13.71      &13.95 & 14.10 &      &max 2444846.36\\
771 &18:22:39.71 +28:30:08.7 &EW   & 0.426878  &13.10 & 13.35 &13.30 &min 2444112.380\\
772 &18:22:48.17 +33:17:26.0 &EA   & 1.55636   &12.75 & 12.95 &12.85 &min 2444847.395&  1\\
773 &18:23:15.75 +29:16:09.3 &EW   & 0.459035  &14.15 & 14.50 &15.45 &min 2443697.316\\
774 &18:23:34.85 +30:50:25.8 &SR:  &17.94      &14.20 & 14.40 &      &max 2443691.4  &  30\\
775 &18:23:39.28 +29:08:08.5 &EW   & 0.464794  &14.90 & 15.15 &15.05 &min 2443669.376\\
776 &18:23:46.16 +28:52:13.6 &RVA  &143.8      &11.75 & 12.30 &12.10 &min 2445197.4  &  31\\

  \noalign{\smallskip}\hline
\end{tabular}
\end{center}
\end{table}
}

\newpage
{\small

\begin{table}
\begin{center}
\caption[]{Table 1(continued)
}\label{Tab:publ-works}

 \begin{tabular}{lllllllll}
  \hline\noalign{\smallskip}
MDV & RA, Dec (J2000.0) & Type & Period& Max& Min &MinII& Epoch& Rem\\
  \hline\noalign{\smallskip}
777 &18:24:01.70 +28:56:09.6 &RRAB & 0.564451  &14.80  &16.20&       &max 2444106.406&  2\\
778 &18:24:07.09 +31:29:23.0 &EA   & 0.691002  &14.30  &14.90& 14.35 &min 2444044.462\\
779 &18:24:11.81 +30:57:18.8 &SR   &56.3:      &16.15  &16.70&       &               &  32\\
780 &18:24:19.23 +29:54:47.4 &EA   & 0.569231  &15.80  &16.50& 15.90 &min 2444109.291\\
781 &18:24:23.19 +30:39:31.0 &RRC  & 0.290012  &15.65  &16.25&       &max 2444077.330\\
782 &18:24:23.55 +32:57:39.5 &EB   & 0.420047  &15.65  &16.20& 15.85 &min 2444877.240\\
783 &18:24:34.68 +29:22:59.8 &RRC  & 0.330565  &13.25  &13.45&       &max 2445197.362&  1\\
784 &18:24:39.29 +30:32:18.7 &HADS & 0.0574812 &14.35  &14.65&       &max 2443672.327&  6\\
785 &18:24:40.47 +32:29:00.0 &EB   & 0.475185  &16.55  &17.10& 16.70 &min 2445224.286\\
786 &18:24:41.04 +33:27:28.2 &RRAB & 0.688098  &15.45  &16.40&       &max 2444081.379\\
787 &18:24:45.24 +30:24:18.4 &LB   &           &16.15  &16.45&       &               &  1\\
788 &18:24:45.73 +32:12:37.1 &LB   &           &14.10  &14.80&       &               &  1\\
789 &18:24:54.51 +33:33:18.0 &EW   & 0.438205  &15.90  &16.30& 16.30 &min 2443282.418\\
790 &18:24:57.45 +33:27:05.0 &EW:  & 0.191276  &16.40  &16.95& 16.90 &min 2443282.418&  33\\
791 &18:24:59.85 +30:07:01.3 &HADS:& 0.0970799 &15.20  &15.40&       &max 2444104.282&  15,7\\
792 &18:25:07.65 +30:41:13.8 &RRC  & 0.294932  &16.50  &17.00&       &max 2444073.419\\
793 &18:25:16.58 +31:29:39.4 &SR   &43.5       &14.35  &14.60&       &max 2443696.4  &  29\\
794 &18:25:23.86 +26:29:42.7 &RRAB & 0.754233  &15.75  &16.25&       &max 2444015.404\\
795 &18:25:29.84 +30:23:59.4 &RRAB & 0.817031  &15.80  &16.45&       &max 2444039.356\\
796 &18:25:35.73 +29:58:26.2 &RRAB & 0.555571  &15.90  &17.00&       &max 2444083.428\\
797 &18:25:36.04 +29:35:21.4 &EW   & 0.644310  &15.65  &16.00& 16.00 &min 2444075.403&  11\\
798 &18:25:37.36 +33:30:08.2 &SR   &38.1       &12.95  &13.15&       &max 2444141.3  &  34\\
799 &18:25:46.82 +27:40:55.7 &RRAB & 0.840954  &14.40  &14.90&       &max 2443748.313\\
800 &18:25:59.08 +27:33:10.0 &HADS & 0.1128665 &15.75  &16.15&       &max 2444073.419\\
801 &18:26:04.44 +35:52:23.6 &EW   & 0.296134  &14.40  &14.60& 14.55 &min 2443253.482\\
802 &18:26:11.38 +33:56:41.9 &LB   &           &14.05  &14.30&       &               &  35\\
803 &18:26:14.10 +32:18:30.3 &EW   & 0.368156  &16.30  &16.75& 16.65 &min 2444875.237\\
804 &18:26:30.30 +27:40:12.3 &RRAB & 0.521100  &15.55  &16.70&       &max 2443757.332\\
805 &18:26:33.06 +33:32:41.6 &EA   & 3.48541   &13.60  &13.85&       &min 2448401.506\\
806 &18:26:35.76 +29:52:32.2 &RRAB & 0.607963  &16.30  &17.20&       &max 2447676.378\\
807 &18:26:55.67 +32:13:25.3 &EW   & 0.419038  &14.20  &14.65& 14.55 &min 2444081.379\\
808 &18:27:04.92 +26:44:34.0 &EW   & 0.399711  &15.35  &15.85& 15.80 &min 2444818.416\\
809 &18:27:10.36 +28:24:08.3 &EA   & 2.00337   &15.25  &15.80&       &min 2444077.330&  36\\
810 &18:27:14.10 +31:26:45.5 &EW   & 0.382948  &14.10  &14.55& 14.50 &min 2443803.254&  1\\
811 &18:27:15.70 +31:00:29.1 &EW   & 0.404429  &15.55  &15.80& 15.75 &min 2443813.217\\
812 &18:27:28.72 +27:35:44.4 &EW   & 0.335950  &14.75  &15.30& 15.30 &min 2444277.240\\
813 &18:27:30.79 +32:57:46.7 &EW   & 0.323209  &15.20  &15.60& 15.50 &min 2445230.283\\
814 &18:27:37.52 +26:52:14.1 &EB   & 0.449438  &14.85  &15.10& 14.95 &min 2444073.419\\
815 &18:27:39.93 +28:14:22.9 &RRC  & 0.298886  &16.80  &17.20&       &max 2444106.406\\
816 &18:27:45.90 +27:54:15.0 &RRC  & 0.289764  &15.80  &16.30&       &max 2444104.282&  7\\
817 &18:27:48.20 +32:13:49.3 &LB   &           &13.65  &13.90&       &               &  37\\
818 &18:27:49.82 +32:40:20.6 &EB   & 0.460073  &15.50  &15.85& 15.70 &min 2444076.345\\
819 &18:27:54.02 +30:47:19.1 &RRC  & 0.394306  &14.60  &15.15&       &max 2443199.553\\
820 &18:27:56.64 +29:13:05.5 &EW   & 0.401254  &15.45  &15.90& 15.90 &min 2444397.450\\
821 &18:27:57.76 +30:51:34.2 &EW   & 0.321032  &16.40  &16.80& 16.70 &min 2443663.483\\
822 &18:28:15.96 +36:26:25.2 &HADS & 0.0465770 &16.05  &16.30&       &max 2444106.406&  2\\

  \noalign{\smallskip}\hline
\end{tabular}
\end{center}
\end{table}
}

\newpage
{\small

\begin{table}
\begin{center}
\caption[]{Table 1(continued)
}\label{Tab:publ-works}

 \begin{tabular}{lllllllll}
  \hline\noalign{\smallskip}
MDV & RA, Dec (J2000.0) & Type & Period& Max& Min &MinII& Epoch& Rem\\
  \hline\noalign{\smallskip}

823 &18:28:19.92 +26:23:11.5 &EA   & 4.10337  & 14.00  &15.00:&      &min 2444846.52 &  1\\
824 &18:28:27.20 +28:44:26.1 &RRAB & 0.590030 & 16.05  &17.05 &      &max 2444821.298\\
825 &18:28:29.35 +33:45:29.0 &EA   & 1.59798  & 14.20  &14.40 &14.25 &min 2444790.328\\
826 &18:28:30.04 +27:38:14.2 &EB   & 0.833553 & 15.05  &15.55 &15.50 &min 2447328.457\\
827 &18:28:43.00 +31:39:04.3 &EW   & 0.393631 & 15.15  &15.35 &15.35 &min 2444164.232&  11\\
828 &18:28:48.47 +29:41:53.9 &EB   & 0.718740 & 15.50  &16.10 &15.80 &min 2448401.506\\
829 &18:29:04.91 +30:52:53.2 &RRC  & 0.345776 & 13.40  &13.55 &      &max 2442869.499\\
830 &18:29:15.55 +32:34:33.6 &E+RS & 9.83730  & 13.30  &13.80 &      &min 2443780.248&  15,38\\
831 &18:29:16.07 +30:50:57.7 &LB   &          & 13.95  &14.30 &      &               &  39\\
832 &18:29:20.58 +31:56:48.1 &SR   &38.5:     & 14.25  &14.55 &      &max 2444839.3  &  40\\
833 &18:29:34.78 +27:07:01.4 &LB   &          & 14.45  &14.75 &      &               &  41\\
834 &18:29:39.86 +33:07:04.3 &EA   & 3.14739  & 15.05  &15.65 &15.10 &min 2444457.310\\
835 &18:29:41.61 +31:17:57.6 &LB   &          & 12.85  &13.20 &      &               &  42\\
836 &18:29:57.15 +29:02:08.3 &RRAB & 0.607354 & 16.25  &16.70 &      &max 2443776.253\\
837 &18:30:09.80 +30:43:07.6 &SR   &67.5      & 13.60  &14.25 &      &max 2444823.3  &  43\\
838 &18:30:20.33 +30:19:23.0 &RRC  & 0.333127 & 16.20  &16.70 &      &max 2444103.301\\
839 &18:30:22.96 +29:06:43.2 &RRAB & 0.603389 & 15.85  &16.95 &      &max 2443696.388\\
840 &18:30:24.56 +31:13:07.6 &EW   & 0.302360 & 16.15  &16.60 &16.55 &min 2443759.336\\
841 &18:30:24.59 +29:28:06.9 &LB   &          & 11.35  &11.75 &      &               &  44\\
842 &18:30:26.69 +27:21:37.9 &SR   &44.5:     & 14.80  &15.05 &      &max 2444039.4  &  29\\
843 &18:30:51.02 +32:08:35.6 &RRC  & 0.348948 & 15.55  &16.10 &      &max 2445251.237\\
844 &18:31:00.79 +30:36:44.4 &EW   & 0.429314 & 15.90  &16.35 &16.25 &min 2444781.425\\
845 &18:31:01.58 +33:26:21.8 &EW   & 0.353845 & 16.15  &16.65 &16.65 &min 2444024.407\\
846 &18:31:14.97 +27:02:49.3 &RRC  & 0.343227 & 13.80  &14.00 &      &max 2443050.290&  1,9\\
847 &18:31:15.83 +34:23:50.4 &EA   & 0.919635 & 15.75  &16.25 &15.85 &min 2444046.393&  2\\
848 &18:31:18.42 +26:41:29.5 &EW   & 0.325537 & 15.10  &15.40 &15.40 &min 2443430.232\\
849 &18:31:22.62 +30:36:44.8 &EB   & 0.843905 & 14.80  &15.10 &15.00 &min 2448095.335\\
850 &18:31:35.22 +27:46:07.5 &EB   & 0.566952 & 15.30  &15.65 &15.45 &min 2443696.421\\
851 &18:31:40.81 +32:58:57.9 &BY:  & 7.540    & 14.15  &14.45 &      &max 2443287.4\\
852 &18:31:48.29 +29:03:21.1 &EA   & 0.952666 & 14.75  &15.20 &14.80 &min 2443814.217\\
853 &18:31:57.99 +29:26:04.4 &LB   &          & 13.65  &14.05 &      &               &  1\\
854 &18:31:59.77 +29:43:38.5 &EA   & 0.542620 & 15.75  &16.30 &15.90 &min 2443814.217\\
855 &18:32:03.89 +32:07:28.3 &RRC  & 0.306341 & 15.55  &16.05 &      &max 2444823.293\\
856 &18:32:06.20 +26:14:47.8 &EB   & 0.540360 & 15.05  &15.60 &15.25 &min 2443691.373&  12\\
857 &18:32:43.73 +28:58:48.3 &EW   & 0.347454 & 15.20  &15.50 &15.50 &min 2444075.377\\
858 &18:32:43.98 +29:56:48.3 &EW   & 0.392832 & 14.60  &15.10 &15.05 &min 2443814.217&  2\\
859 &18:32:53.51 +30:11:23.3 &RRAB & 0.592867 & 14.60  &15.55 &      &max 2443814.217&  2\\
860 &18:33:15.64 +30:58:17.4 &LB   &          & 13.65  &13.90 &      &               &  15\\
861 &18:33:17.48 +31:32:57.6 &EW   & 0.330509 & 13.90  &14.20 &14.15 &min 2443748.313&  1\\
862 &18:33:31.83 +28:16:18.4 &HADS:& 0.131393:& 16.50  &17.10 &      &max 2444025.341&  12\\
863 &18:33:46.78 +29:50:40.1 &EW   & 0.343149 & 15.20  &15.40 &15.40 &min 2443718.385&  45\\
864 &18:33:47.82 +31:47:47.6 &CWA: &11.37     & 16.15  &16.45 &      &max 2443699.5\\
865 &18:34:00.46 +28:59:07.1 &EW   & 0.364373 & 15.90  &16.40 &16.35 &min 2443672.327\\
866 &18:34:29.58 +28:41:27.7 &RRC: & 0.369664 & 14.85  &15.05 &      &max 2444489.313&  7\\
867 &18:34:51.12 +34:00:20.3 &EA   & 2.72607  & 16.10  &17.05 &16.20 &min 2443701.418\\
868 &18:35:04.26 +32:25:13.5 &EW   & 0.375475 & 15.75  &16.05 &16.00 &min 2443722.368\\
869 &18:35:26.63 +29:39:14.4 &EW   & 0.502887 & 15.10  &15.90 &15.80 &min 2443783.253\\
870 &18:36:56.84 +34:05:57.2 &RRC  & 0.331094 & 14.00  &14.70 &      &max 2443190.574\\

  \noalign{\smallskip}\hline
\end{tabular}
\end{center}
\end{table}
}

Remarks to Table 1

1. Varies in NSVS data. 2. Varies in Catalina data. 3. From NSVS
data, $P = 48.6^d$. 4. A small-amplitude variable from NSVS data,
$P = 56.6^d$ not detected. 5. A small-amplitude variable from NSVS
data, $P = 5.846^d$. 6. Our CCD observations confirm variability,
type, period. 7. Also possible is type EW with a twice longer
period. 8. MinII--MinI = $0.58^P$. 9. Double star. 10. A twice
shorter period is possible. 11. Also possible is type RRC with a
twice shorter period. 12. Our CCD observations confirm
variability. 13. The faint component of the pair varies,
photometry from scans merges the two components. The magnitudes in
Table 1 are from eye estimates for the faint component. The range
from scans for combined brightness is $14.05^m - 14.40^m$. 14.
Period varies? 15. Varies in ASAS-SN data. 16. NSVS data suggest
$P = 33.1^d$. 17. Catalina data suggest $P = 0.359270^d$. 18. NSVS
data suggest $P = 17.0^d$. 19. NSVS data suggest $P = 37.7:^d$.
20. NSVS data suggest $P = 172^d$. 21. NSVS data suggest $P =
77.3:^d$. 22. NSVS data suggest $P = 54.3:^d$. 23. $P =
0.236266^d$ is also possible. 24. Type EB with a twice longer
period is not excluded. 25. Blazhko effect? 26. NSVS data suggest
$P = 48.8^d$. 27. NSVS data suggest $P = 33.9^d$. 28. $P =
0.412568^d$ is also possible. 29. Varies in NSVS data, no period.
30. From NSVS data, small amplitude, $P = 17.25^d$. 31. NSVS data
suggest $P = 143.8^d$. 32. NSVS data suggest $P = 53.4^d$. 33.
Also possible is type HADS with a twice shorter period. 34. NSVS
data suggest $P = 38.3^d$. 35. NSVS data suggest type SR and $P =
34.1^d$. 36. A twice longer period is possible. 37. NSVS data
suggest type SR and $P = 37.1^d$. 38. X-ray source 1RXS
J182915.3+323440. 39. Emission-line star StH$\alpha$ 151, M0e. 40.
NSVS data suggest $P = 43.2^d$. 41. ASAS-SN data suggest type SR
and $P = 47.4^d$. 42. ASAS-SN data suggest type SR: and $P =
35.9^d$. 43. NSVS data suggest $P = 34.6^d$. 44. NSVS data suggest
type SR: and $P = 49:^d$. 45. $P = 0.292892^d$ is also possible.

\begin{figure}
   \centering
   \includegraphics[width=15cm, angle=0]{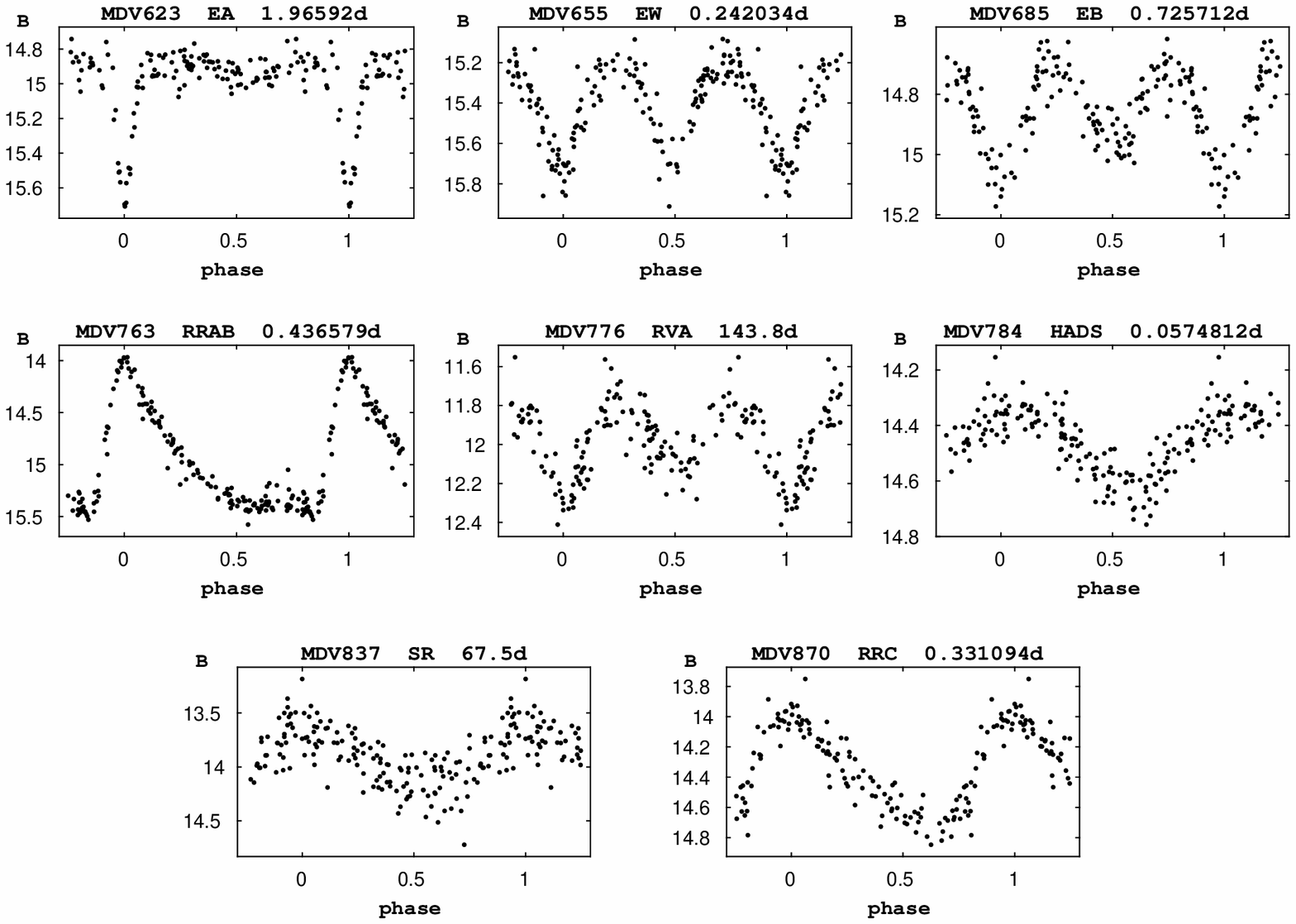}
   \caption{Sample photographic light curves for some of the new variable stars of different types.}
   \label{Fig1}
   \end{figure}

\begin{figure}
   \centering
   \includegraphics[width=11cm, angle=0]{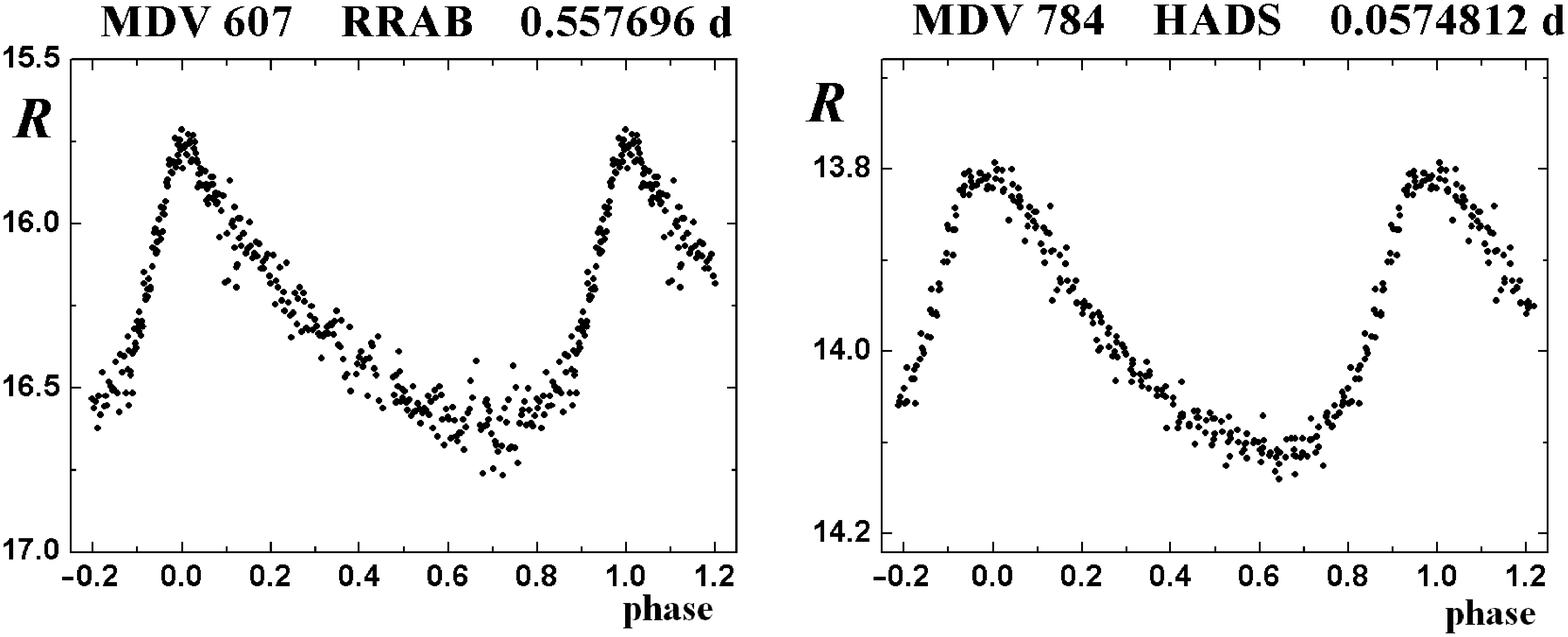}
   \caption{CCD light curves for two MDV variables.}
   \label{Fig2}
   \end{figure}

\section{Automated Classification of the New MDV Stars}

Several teams are currently developing algorithms permitting to
perform automated classification of variable stars from
photometric data. Such algorithms are of extreme importance for
future work on variables stars discovered in large-scale projects,
like expected discoveries of the GAIA space mission. Among the
existing algorithms, the majority have been tested only on
high-quality photometric data. Our noisy light curves are a real
challenge for such programs.

By its nature light curves are not structured since they are
noisy, the number of observations per object depends on the
observing strategy and have missing values. An additional
preprocessing must be done by computing statistical descriptors or
features. Their mission is to reduce dimensionality and make the
data uniform to be handled by different algorithms.

Machine learning can be divided into three broad categories:
supervised learning, which uses labelled data to learn patterns in
each class; reinforcement learning, which interacts with the
environment such as sensors or a human; and unsupervised learning
uses just the data to learn patterns, which later can be used to
make visualizations or as inputs for another algorithm.

In their seminal work, Debosscher et al. (2007) constructed a set
of 28 features. They were based on analytical fits and periods
obtained using the Lomb--Scargle periodogram. They tested it on a
Mixture of Gaussians, Neural Networks and Support Vector Machines
(SVM, Cortes and Vapnik 1995), obtaining an average precision of
70\%. The work of Kim et al. (2011) found that Random Forest (RF,
Breiman 2001) outperforms SVM using 11 features. Richards et al.
(2011) obtained similar conclusions using 53 features. Pichara et
al. (2012) used a database that included autoregressive features
for training an RF, improving previous results. The model was
enhanced later by Pichara and Protopapas (2013), who used
graphical models to fill missing data keeping the computational
cost the same.

Since features do not guarantee proper classification, Mackenzie
et al. (2016) applied a different approach, developing the first
unsupervised feature learning algorithm. It creates a dictionary
of words with fragments of the light curves to make a new
representation. With them, an SVM classifier was trained,
obtaining similar and even better results than the RF while
reducing the computational cost significantly.

The initial light curves of the variables in the field of 104 Her
were subjected to automated classification based on the RF
algorithm since it proved to be the most effective one.

We used the FATS features python package (Nun et al. 2015) using
only time and magnitude fields, as the data does not have
uncertainty estimates for each individual measurement.
Additionally, they were augmented with 2MASS (Skrutskie et al.
2006) and WISE (Wright et al. 2010) colors.

To train the classifier, we used stars earlier announced in the
MDV program (Kolesnikova et al. 2010; Sokolovsky et al. 2014). For
different reasons, we rejected 18~objects from these lists. In
particular, the less represented classes were removed as the RF
algorithms cannot generalize a class with a single example. The
resulting classes used for training were: EW (129 stars); LB
(115); RRAB (99); EB (62); RRC (60); SR (55); EA (45); HADS (12),
the total of 577~stars.

The result was expressed as a table which listed the three top
probabilities of classification, from 0\% to 100\%, for each
variable star. We compared them to the results of our
non-automated classification presented in Table~1. The comparison
is based on 273 stars; two stars were added to our list of
discoveries later.

For 155 stars (57\%), the classification in Table 1 coincides with
the most probable suggestion of the automated classification. The
classification for 85 stars (31\%) is that suggested automatically
as the second or the third option. The non-automated
classification disagrees with any option from the automated
classification in 12\% of all cases.

A frequent case of non-automated classification agreeing with the
second or third automated option is that a different type of
eclipsing variables is selected. Distinguishing between the EA,
EB, EW types is sometimes really somewhat tricky. In our opinion,
such cases should also be considered a relative success of the
automated technique.

From the stars the automated classification was least successful
(12\%), we find that 9 stars ($\sim$3\%) corresponded to classes
that were not represented in the training set, which are usually
the most interesting ones. The remaining 24 stars ($\sim$9\%) were
misclassified even when the class was represented. For noisy light
curves of RR Lyrae variables and most of the HADS stars, the
classifier tends to decide for the eclipsing types. The SR type
was often confused with LB stars. We find that the features cannot
describe irregularity and semi-regularity as different object
classes.

Some interesting cases could not be solved automatically. Thus,
the beautiful RV Tauri star MDV776 got the following type
suggestions automatically: SR (probability 56\%), LB (probability
43\%), and EA (probability 0\%). Variable- star astronomers know
that variations of RV Tauri stars are indeed semiregular and that
they such stars use to show minima sharper than maxima.

We hope that further improvement of the algorithm will prevent it
from missing stars that are most attractive to astrophysicists.
Thus, the automated classification can provide a correct
classification for a majority of all stars. Noisy data make the
task more difficult (our first attempts used ``cleaned'' light
curves and were more successful). To provide an algorithm
completely satisfying compilers of variable-star catalogs,
additional work on the code and its application to noisy data are
needed.

One of the possible solutions is to enhance the dataset with
OGLE-III $V$-band variables and gradually retrain the model as new
stars are detected and classified.

\section{Conclusions}

We continue our work on digitizing photographic plates of the
Moscow collection and on finding new variable stars using
digitized plates. In this paper, we have presented our results for
the $10^\circ\times10^\circ$ field centered at the star 104 Her.

In total, we discovered 275 new variable stars in the field,
mainly eclipsing binaries but also pulsating stars and, possibly,
rotating spotted stars. This new study generally confirms our
earlier finding that HADS variables are much better represented
among the new discoveries than in the General Catalogue of
Variable Stars.

The special feature of this study that we have made an attempt of
automated classification of the discovered variable stars based on
our photographic photometry. The results were compared to our
traditional classification. In 88\% of all cases, we were able to
achieve good or satisfactory results; the automated classification
failed in 12\% of cases.

\begin{acknowledgements}

The authors would like to thank the referee for very valuable
suggestions that helped us to improve the manuscript.

\end{acknowledgements}

\label{lastpage}


\begin{thebibliography}{99}

\bibitem[2005]{bacher}Bacher A., Kimeswenger S., Teutsch P., 2005, \mnras, 362, 542

\bibitem[2016]{bal}Balona L. A., 2016, \mnras, 459, 1097

\bibitem[1996]{bertin}Bertin E., Arnouts S., 1996, \aaps,
117, 393

\bibitem[2001]{brei}Breiman L., 2001, Machine Learning, 45, 5

\bibitem[1995]{corvap}Cortes C., Vapnik V., 1995, Machine
Learning, 20, 273

\bibitem[2007]{debosh}Debosscher J., Sarro L.M., Aerts C.,
Cuypers J., Vandenbussche B., Garrido R., Solano E., 2007, \aap,
475, 1159

\bibitem[2009]{drake}Drake A. J., Djorgovski S. G., Mahabal A., Beshore E.,
Larson S., Graham M. J., Williams A., Christensen E., Catellan M.,
Boattini A., Gibbs A., Hill A., Kowalski R., 2009, \apj, 696, 870

\bibitem[2008]{hogg}Hogg D. W., Blanton M., Lang D., Mierle K., Roweis S.,
2008, ASPC, 394, 27

\bibitem[2009]{harv}Grindlay J., Tang S., Simcoe R., Laycock
S., Los E., Mink D., Doane A., Champine G., 2009, ASPC, 410, 101

\bibitem[2011]{kim}Kim D.-W., Protopapas P., Byun Y.-Ik,
Alcock Ch., Khardon R., Trichas M., 2011, \apj, 735, article id.
68

\bibitem[2017]{kochan} Kochanek C. S., Shappee B. J., Stanek K. Z., Holoien T. W.-S.,
Thompson T. A., Prieto J. L., Dong S., Shields J. V., Will D.,
Britt S., Perzanowski, D. Pojmanski G., 2017, \pasp, 129, 104502

\bibitem[2008]{kol08}Kolesnikova D. M., Sat L. A., Sokolovsky K. V., Antipin S. V.,
Samus N. N., 2008, Acta Astronomica, 58, 279

\bibitem[2010]{kol10}Kolesnikova D. M., Sat L. A., Sokolovsky K. V., Antipin S. V.,
Belinskii A. A., Samus N. N., 2010, Astronomy Reports, 54, 1000

\bibitem[2010]{lang}Lang D., Hogg D. W., Mierle K., Blanton M., Roweis S.,
2010, \aj, 139, 1782

\bibitem[2016]{mack}Mackenzie C., Pichara K., Protopapas P., 2016, \apj, 820, article id. 138

\bibitem[2003]{monet}Monet D. G., Levine S. E., Canzian B., Ables H. D., Bird
A. R., Dahn C. C., Guetter H. H., Harris H. C., Henden A. A.,
Leggett S. K., Levison H. F., Luginbuhl C. B., Martini J., Monet
A. K. B., Munn J. A., Pier J. R., Rhodes A. R., Riepe B., Sell S.,
Stone R. C., Vrba F. J., Walker R. L., Westerhout G., Brucato R.
J., Reid I. N., Schoening W., Hartley M., Read M. A., Tritton S.
B., 2003, \aj, 125, 984

\bibitem[2015]{nun} Nun I., Protopapas P., Sim B., Zhu M., Dave R., Castro N.,
Pichara K., 2015, ArXiv 1506.00010

\bibitem[2013]{pp13}Pichara K., Protopapas P., 2013, \apj,
777, article id. 83

\bibitem[2012]{ppkmt}Pichara K., Protopapas P., Kim D.-W.,
Marquette J.-B., Tisserand P., 2012, \mnras, 427, 1284

\bibitem[2011]{riea}Richards J.W., Starr D.L., Butler N.R.,
Bloom J.S., Brewer J.M., Crellin-Quick A., Higgins J., Kennedy R.,
Rischard M., 2011, \apj, 733, article id. 10

\bibitem[2017]{sam}Samus N. N., Kazarovets E. V., Durlevich O. V., Kireeva N. N.,
Pastukhova E. N., 2017, Astronomy Reports, 61, 80

\bibitem[1999]{shug}Shugarov S., Antipin S., Samus N., Danilkina T., 1999, Acta
Historica Astronomiae, 6, 81

\bibitem[2006]{2mass}Skrutskie M. F., Cutri R. M.,  Stiening R. et
al., 2006, \aj, 131, 1163

\bibitem[2014]{sok1}Sokolovsky K. V., Antipin S. V., Zubareva A. M., Kolesnikova
D. M., Lebedev A. A., Samus N. N., Sat L. A., 2014, Astronomy
Reports, 58, 319

\bibitem[2017]{sok2}Sokolovsky K. V., Lebedev A. A., 2017, arXiv: 1702.07715

\bibitem[2017]{woz1}Wo\'zniak P. R., Vestrand W. T., Akerlof C. W., Balsano R., Bloch J., Casperson
D., Fletcher S., Gisler G., Kehoe R., Kinemuchi K., Lee B. C.,
Marshall S., McGowan K. E., McKay T. A., Rykoff E. S., Smith D.
A., Szymanski G., Wren J., 2004a, \aj, 127, 2436

\bibitem[2017]{woz2}Wo\'zniak P. R., Williams P. R., Vestrand
W. T., Gupta V., 2004b, \aj, 128, 2965

\bibitem[2010]{wise}Wright E. L., Eisenhardt P. R. M., Mainzer A. K. et al., 2010, \aj, 140, 1868

\end{thebibliography}
\end{document}